# Pulsed Laser Deposition of epitaxial titanium diboride thin films

V.Ferrando[a], D.Marré[a], P.Manfrinetti[b], I.Pallecchi[a], C.Tarantini[a], C.Ferdeghini[a]

[a] INFM-LAMIA, Dipartimento di Fisica, Via Dodecaneso 33, 16146 Genova, Italy

[b] INFM, Dipartimento di Chimica e Chimica Industriale, Via Dodecaneso 31,16146 Genova, Italy

**Abstract**

Epitaxial titanium diboride thin films have been deposited on sapphire substrates by Pulsed Laser Ablation technique. Structural properties of the films have been studied during the growth by Reflection High Energy Electron Diffraction (RHEED) and ex-situ by means of X-ray diffraction techniques; both kinds of measurements indicate a good crystallographic orientation of the $TiB_2$ film both in plane and along the *c* axis. A flat surface has been observed by Atomic Force Microscopy imaging. Electrical resistivity at room temperature resulted to be five times higher than the value reported for single crystals. The films resulted to be also very stable at high temperature, which is very promising for using this material as a buffer layer in the growth of magnesium diboride thin films.

## 1. Introduction

Diborides are a well known class of materials that have been used in a number of applications since the Fifties. In particular $TiB_2$ presents very high hardness (up to 3000 HV), high melting point (3220 °C), good wear and corrosion resistance and metallic properties such as low friction coefficient and high thermal and electrical conductivity. These properties make this material suitable for applications as high performance cutting and forming tools, highly corrosion resistant coatings, thermal barriers in fusion and chemical reactors and diffusion barriers on Si in semiconductors technology [1]. For these applications, the material is produced in form of thin or thick films by a variety of deposition techniques [2-4] (sputtering, chemical vapor deposition, laser ablation etc), but epitaxy is not required. The discovery of superconductivity in $MgB_2$ [5] renewed the interest in this material and, in general, in the entire class of diborides both for applications and fundamental studies. In particular, the deposition of diborides in form of epitaxial thin films could be very promising because they can be used as buffer layer to improve quality of $MgB_2$ thin films. In fact, despite the great efforts done, epitaxial growth of $MgB_2$ thin films is rarely reported in literature [6-9]; therefore an isostructural buffer layer can help for this purpose. For the fabrication of heterostructures in fact, epitaxial thin films with flat surface are desirable. Moreover, ref. [10] demonstrates that variation in the lattice parameter *a*, in this case produced by the tensile strain coming from boron crystal substrate, can vary the $MgB_2$ critical temperature. In particular, an increase of lattice parameters can produce a slight but significant $T_C$ enhancement. It must be noted that among $AlB_2$-type isostructural diborides, compounds either with larger *a* parameter (3.1694 and 3.1478 Å for $ZrB_2$ and $ScB_2$ respectively) or smaller (3.038 Å for $TiB_2$) are present [11]. Therefore, by an appropriate choice of the buffer layer, in case of epitaxial growth of magnesium diboride, a tuning of $T_C$ can be obtained. Finally, it must also be noted that other diborides do not suffer of the volatility of one element and therefore more usual deposition conditions can be adopted. In this paper our results in the deposition of high quality $TiB_2$ thin films using Pulsed Laser Ablation technique are presented.

## 2. Experimental

The Pulsed Laser Ablation deposition apparatus consists of an ultra high vacuum chamber equipped with a RHEED in which a pressure down to $10^{-10}$ mbar can be reached and of a KrF excimer laser (wavelength = 248 nm). Details on the experimental setup are presented in ref. [12]. Films were grown starting from a dense stoichiometric commercial $TiB_2$ target (MTI Components) with a diameter of 2 cm. Different substrates, i.e.: 6H-SiC, *c*-cut $Al_2O_3$ and MgO (111) have been utilized. SiC and $Al_2O_3$ substrates have the same hexagonal structure of $TiB_2$ but with a large lattice mismatch. MgO, cut in the (111) direction, has the same surface symmetry but with a smaller lattice mismatch (around 3%). Sapphire and magnesium oxide are also less expensive with respect to silicon carbide which has already been used for the preparation of $TiB_2$ thin films [13].

The deposition was performed in a background pressure of $10^{-9}$ mbar, the laser frequency was varied in the range 3-10 Hz and the beam fluency was around 2-5 $J/cm^2$. The substrate temperature has been changed between 600 and 800 °C; after deposition the films were either immediately quenched to room temperature or slowly cooled to room temperature (half an hour). The growth rate, evaluated performing reflectivity measurements, resulted to be very low, around 0.002 Å per laser shot. An example of these measurements is showed in fig.1; oscillations in the diffracted intensity correspond to a thickness of about 240 Å and are a clear evidence of a very smooth surface. In situ RHEED analysis have been carried out in order to verify growth mode and in-plane orientation of the film. We used an ER 2035 K RHEED system by Staib Instrumente at 20 kV of beam energy and with an incidence angle of ~ 1°. The RHEED diffraction patterns observed on the screen are registered by a CCD camera and then analyzed by a computer.

## 3. Results and discussion

Our best results have been obtained for samples grown on sapphire substrates at a temperature of 720 °C, followed by a quenching to room temperature immediately after the growth.

For samples so prepared, bright diffraction spots are clearly evident at all stages of deposition, up to film thickness of 240 Å; this result indicates an ordered growth of $TiB_2$ along with a good quality of the film surface. In fig. 2, three RHEED images are reported: in figs. 2a and 2b the clean $Al_2O_3$ substrate and a 20 Å thick film are shown. The transmission pattern of fig. 2b indicates that the titanium diboride grows in a three-dimensional mode. By rotating the sample with respect to the electron beam direction, RHEED patterns with different spacing appear. In particular, starting from pattern of fig. 2b, when the film is rotated of 30° another family of spots becomes visible (fig. 2c), its spacing being higher than in the previous case. The observed ratio between these two spacing appears to be exactly $\sqrt{3}$, which corresponds to the value expected for a rotation of 30° around the *c* axis of the $AlB_2$-type hexagonal structure. Moreover, the ratio between spacing of reciprocal lattice of the substrate and that of the growing film resulted to be 1.56, which is exactly consistent with the ratio between the *a* parameter of their respective unit cells. This analysis clearly shows that, in the above conditions, $TiB_2$ can grow epitaxially on sapphire. Furthermore, from the comparison between families of patterns belonging to the film with those of the substrate, we could conclude that growth of TiB2 films on sapphire is a 'hexagon-on-hexagon' growth, without any lattice rotation.

Figure 3 shows a typical X-ray diffraction pattern, in a logarithmic scale, of a 100 Å thick $TiB_2$ film, grown on sapphire. A part from peaks of the Al sample holder at 38.4° and 44.7°, only reflections (00l) of titanium diboride have been detected, thus confirming that the film is grown with the $c$ axis perpendicular to the substrate. The broadness of the (00l) peaks can suggest either a nanometric grain size along $c$ axis direction or the presence of stress in the lattice. The $c$ parameter, calculated from the angular position of these peaks, resulted to be 3.270 ± 0.005 Å, a value slightly higher than the one given for bulk (3.239 Å [11]). The very good $c$ axis orientation of the film is evident also from its rocking curve measurement around the (001) reflection (fig. 4). The very low value of FWHM (0.15°) indicates a good alignment of the crystallographic planes along the $c$ axis.

By changing deposition parameters, or substrate, such a good structural properties are suddenly lost. Indeed, films grown on $Al_2O_3$ substrate below 700 °C or above 750 °C show poorer intensity of reflections; a slow cooling rate seems also not to improve the crystallization of the $TiB_2$ phase.

Films grown on MgO do not show a similar good orientation and surprisingly, we were not able to deposit $TiB_2$ on SiC, differently from what reported in ref. [13].

The use of $TiB_2$ as a buffer layer to improve $MgB_2$ deposition requires a good thermal stability at high temperature. In fact, due to the volatility of Mg, different two-step techniques have been developed for thin films growth [6,7,9,10]; usually, a low temperature precursors deposition followed by an annealing at high temperature (800-900 ºC in a Mg atmosphere) is used. To verify the temperature stability of $TiB_2$ films we have prepared, a sample, wrapped in a zirconium foil and sealed in a Ta crucible under Ar, has been annealed at 850 °C for one hour under vacuum. After annealing, a second X-ray analysis has been performed to check for any change in the structural quality of the film. Only a sensible shift in the $c$ parameter (3.220 ± 0.006 Å), towards a value closer to that of bulk has been observed, indicating a relaxation of the lattice strain. The fact that relaxation does not reflect in a narrowing of the peaks, suggests that their broadness can be ascribed to a nanostructuration along the $c$ axis direction rather than to a lattice strain.

In order to epitaxially grow $MgB_2$ on $TiB_2$, a very good surface morphology is also necessary. An Atomic Force Microscopy (AFM) image of a film, recorded on an area of 1.2 µm x 12 µm, is shown in fig.5. Although some particulates due to the deposition technique are still present, the surface appears very smooth, with a roughness of about 10 Å, low enough to use the film as a buffer layer.

From an electrical point of view, the resistivity of our films, measured by a standard four-probe method, is of the order of 30 µΩcm at room temperature; this value is five times higher than that of single crystals (6.6 µΩcm) [14], differently from what previously reported in ref [13] for thin films, where a resistivity very close to this value (8 µΩcm) have been found.

## 3. Conclusions

Epitaxial thin films of $TiB_2$ have been deposited on *c*-cut $Al_2O_3$ substrates by Pulsed Laser Ablation technique. The structural properties have been studied by RHEED and X-ray diffraction techniques. The films resulted to be both in-plane and *c* axis oriented; moreover, they presented a flat surface, as observed by AFM imaging. Electrical resistivity at room temperature resulted to be five times higher than that of single crystals. Finally, the films so obtained resulted to be stable at high temperature, a result that is very promising for using this material as a buffer layer for $MgB_2$ thin films growth.

**Figure Captions**

Figure 1. X-ray reflectivity measurement: the oscillations in the diffracted intensity indicate a thickness of about 30 nm and are a clear evidence of a very smooth surface.

Figure 2. RHEED images of: a) the clean $Al_2O_3$ substrate, b) the film after 20 Å and c) the film rotated of 30°.

Figure 3. X-Ray diffraction pattern for a 10 nm $TiB_2$ thin film grown on $Al_2O_3$ c-cut substrate.

Figure 4. Rocking curve around the 001 reflection of the same film of fig.3. The FWHM is 0.15º.

Figure 5. AFM image of a 10 nm $TiB_2$ thin film grown on $Al_2O_3$ substrate recorded on an area of 1.2μm x 1.2μm

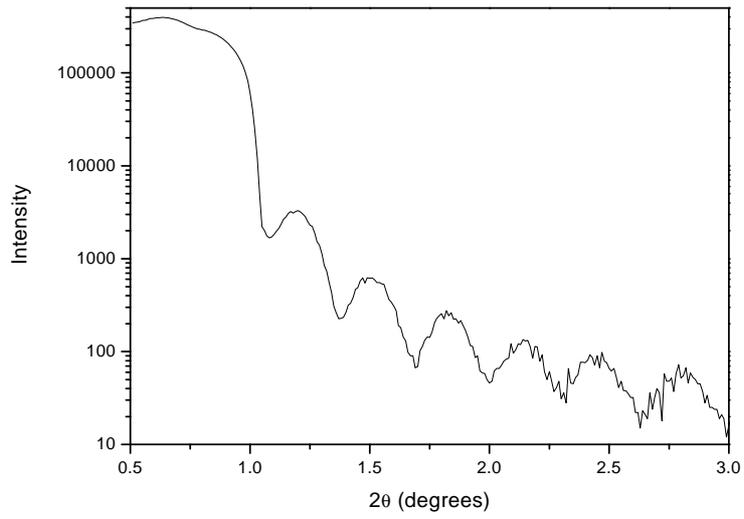

**Figure 1**

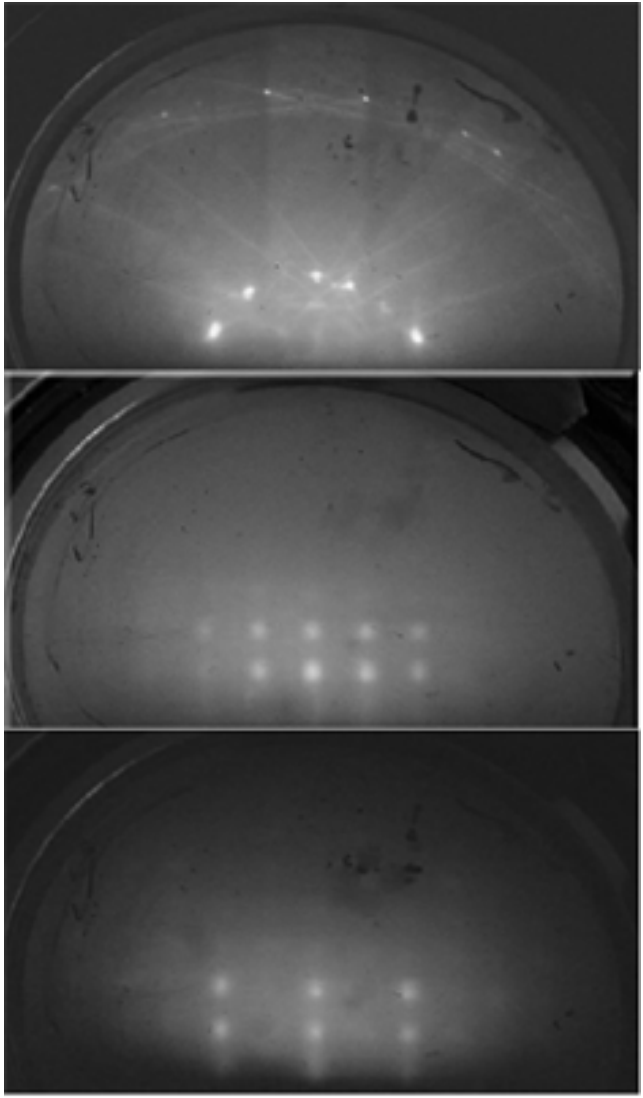

**Figure 2**

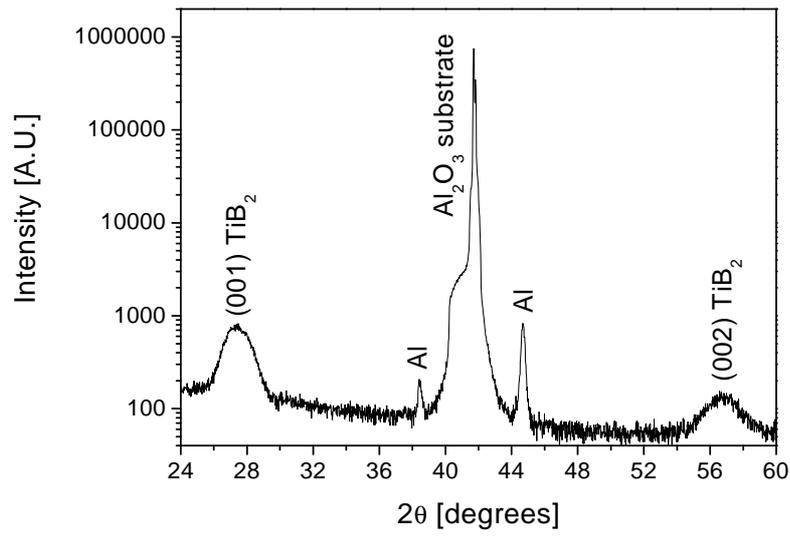

**Figure 3**

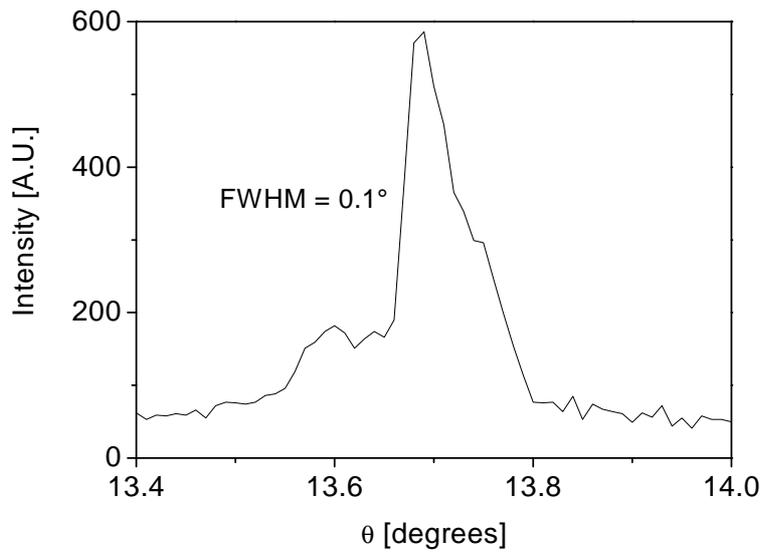

**Figure 4**

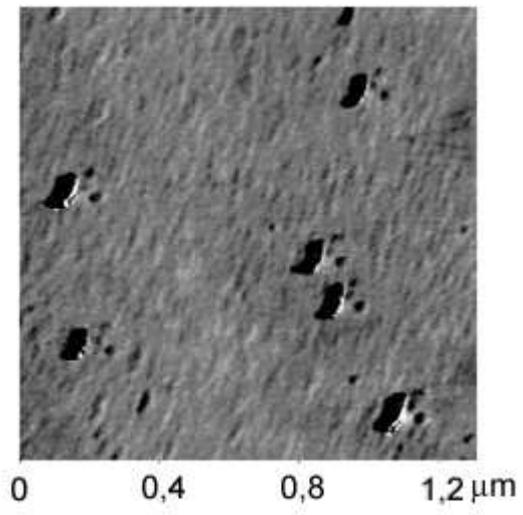

**Figure 5**